\documentclass[]{spie}  

\usepackage{amsmath,amsfonts,amssymb}
\usepackage{graphicx}
\usepackage[colorlinks=true, allcolors=blue]{hyperref}

\title{Unified Multi-scale Feature Abstraction for Medical Image Segmentation}

\author[a]{Xi Fang}
\author[b]{Bo Du}
\author[c]{Sheng Xu}
\author[c]{Bradford J. Wood}
\author[a]{Pingkun Yan}
\affil[a]{Rensselaer Polytechnic Institute, Troy, United States}
\affil[b]{Wuhan University, Wuhan, China}
\affil[c]{National Institutes of Health, Bethesda, United States}

\begin{document} 
\maketitle



\section{Description of purpose}
\label{sec:intro}  

Automatic medical image segmentation, an essential component of medical image analysis, plays an important role in computer-aided diagnosis. For example, locating and segmenting the liver can be very helpful in liver cancer diagnosis and treatment. The state-of-the-art models in medical image segmentation are variants of the encoder-decoder architecture such as fully convolutional network (FCN) and U-Net \cite{unet}. A major focus of the FCN based segmentation methods has been on network structure engineering by incorporating the latest CNN structures such as ResNet \cite{He2016DeepRL} and DenseNet \cite{Huang2017DenselyCC}. In addition to exploring new network structures for efficiently abstracting high level features, incorporating structures for multi-scale image feature extraction in FCN has helped to improve performance in segmentation tasks. In this paper, we design a new multi-scale network architecture, which takes multi-scale inputs with dedicated convolutional paths to efficiently combine features from different scales to better utilize the hierarchical information.

\section{METHODS}

The proposed MIMO-FAN first performs multi-scale analysis to the input image by using spatial pyramid pooling to obtain scene context information. After the first level convolutional blocks with shared kernels, image-level contextual features that interpret the overall scene can be extracted from these inputs in different scales. Starting from there, a notable feature of MIMO-FAN is that features to be fused at a certain level all go through the same number of convolutional layers using dense cross-scale connections (DCCs), which help to keep the hierarchical structure for better abstraction. Each DCC module employs residual connections for the convolutional blocks and dense connections \cite{Huang2017DenselyCC} between different scale features at the same depth. Unlike the classical U-net based methods, where the scale only reduces when the convolutional depth increases, MIMO-FAN has multi-scale features at each depth and therefore both global and local context information can be fully integrated to augment the extracted features. Furthermore, inspired by the work of deep supervision, we further introduce deep pyramid supervision (DPS) to the decoding side for generating and supervising outputs of different scales, which helps to alleviate the gradient vanishing problem and generate good segmentation masks at different scales. Finally, the two largest probability maps are fused together to achieve a more reliable segmentation by scale fusing (SF).
Details of the DCC and DPS modules are provided as follows.

\subsection{Dense Cross-scale Connections (DCCs)}
It has been shown that global feature extraction and contextual integration are beneficial for semantic segmentation.
Instead of extracting multi-scale features at the very late stage of the convolutional networks, we propose to obtain multi-scale features from the beginning of the network to preserve context information of input images and utilize features of different scales during the entire network.
At a certain level, the smaller the scales are, the more global context information that the features may contain. Thus, to efficiently augment feature representation ability, we develop DCC blocks with a new skip connection to fuse feature maps in different scales at the same level, as shown in Fig.~\ref{fig:Concave}(B). We add dense connections between different scale features at the same level. Through DCCs, features in different scales are fused and reused, which makes the features more representative and contain more hierarchical information. In the encoder part, MIMO-FAN uses top-down connections to combine multi-scale feature maps, while the bottom-up order is used in the decoder part to gradually decode high level features.\\

\subsection{Deep Pyramid Supervision (DPS)}

\begin{figure}[t]   
	\center{\includegraphics[width = \textwidth] {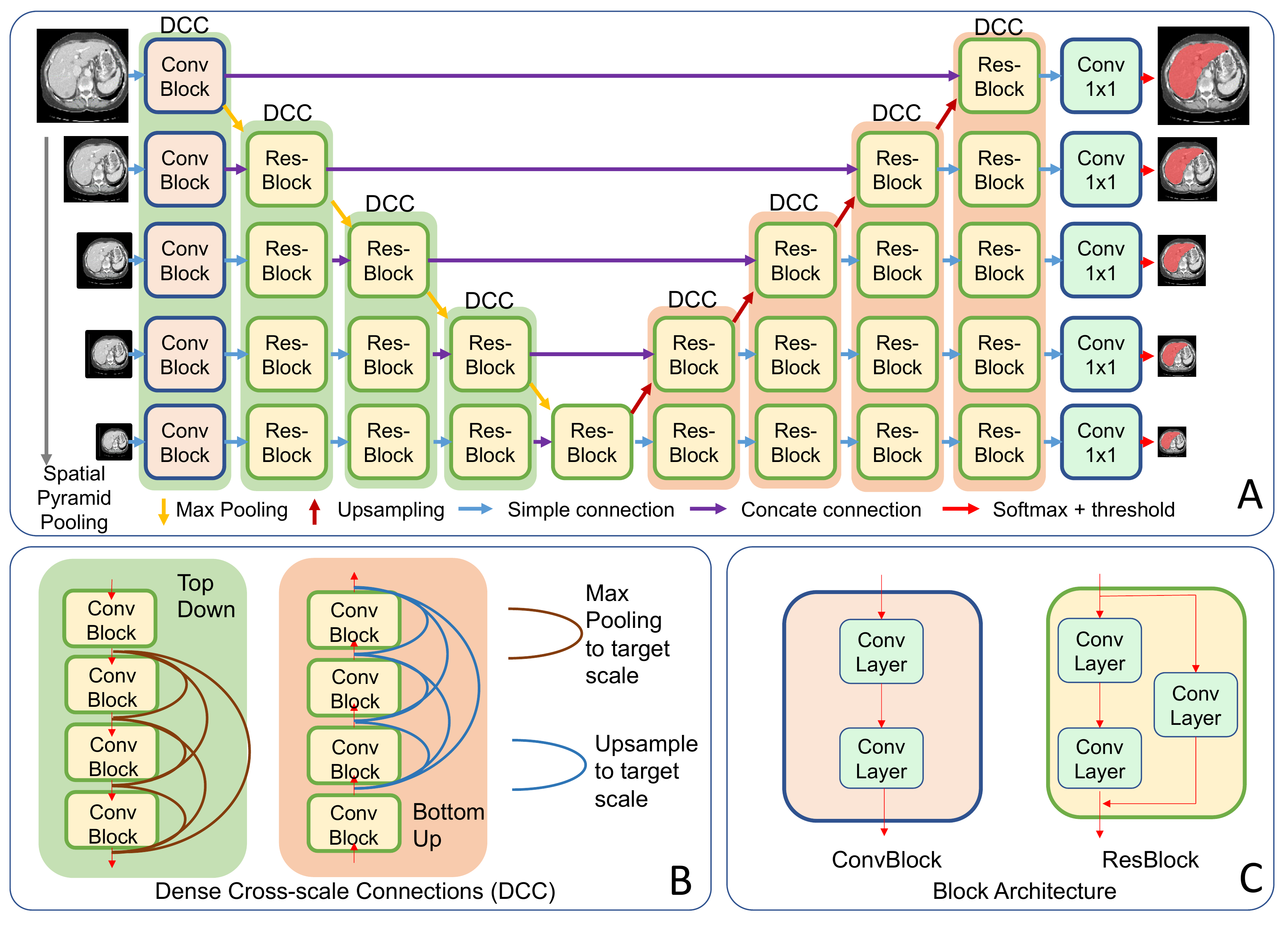}}        
	\caption{Overview of the MIMO-FAN architecture. (A) Information propagation from multi-scale input to hierarchical combination of same level features through dense cross-scale connection (DCC) in (B). (C) Detailed structure of the network blocks.}  
	\label{fig:Concave}
\end{figure}

To enforce efficient feature abstraction at small scales and deep levels, we propose deep pyramid supervision (DPS) for supervising outputs at various scales. To deal with the variation of output sizes, we perform the spatial pyramid pooling operation to the ground truth segmentation to generate labels in all output scales. The training loss is computed by using the output and ground truth segmentation at the same scale. Weighted cross entropy is used as the loss function in our work, which is defined as
\begin{equation}
L = -\frac{1}{S}\sum_{s=1}^{S}\sum_{c=0}^{1}w_{i,s}^c y_{i,s}^c \log p_{i,s}^c,
\end{equation}
where $p_{i,s}^c$ denotes the predicted probability of voxel $i$ belonging to class $c$ (background or liver) in scale $s$, $y_{i,s}^c$ is the ground truth label in scale $s$, $N_s$ denotes the number of voxels in the scale $s$, and $w_i^c$ is weighting parameter for different classes. Empirically, we set the weights to be 0.2 for background, 1.2 for liver. The total number of scales $S$ is set to be 5 in our work corresponding to the illustration in Fig.~\ref{fig:Concave}(A). 

\section{RESULTS}

We extensively evaluated our model on LiTS (Liver tumor segmentation challenge) datasets, which are composed of 131 training and 70 test datasets. The data were collected from different hospitals and the resolution of the CT scans vary between 0.45mm and 6mm for intra-slice and between 0.6mm and 1.0mm for inter-slices, respectively. The size of each slices is 512$\times$512 pixels. To speed up the model training, we resized the axial slices into 256$\times$256 pixels, where the boundary information is still well preserved. Five-fold cross validation was employed to evaluate the performance of the models on the challenge training datasets. When preparing the test result submission to the challenge, we use majority voting to combine the outputs of the five models to get the final segmentation. Our implementation code will be open-sourced once the paper is accepted.

\subsection{Comparison with other methods}

\setlength{\tabcolsep}{6pt}

Most of the state-of-the-art methods on liver CT image segmentation takes two steps to complete the task, where a coarse segmentation is used to locate the liver and followed by fine segmentation step to obtain the final segmentation \cite{han_automatic_2017,8379359}. Many works combine the 2D and 3D features together to improve the segmentation performance \cite{8379359}. However, those methods can be computationally expensive. For example, the method of Li et al. \cite{8379359} takes 21 hours to train the 2D DenseUNet and another 9 hours to finetune the H-DenseUNet with two Titan Xp GPUs. By constrast, our proposed method completes the training of one model on a single Titan Xp GPU in 3 hours. In addition, our 2D network segments the liver in one step and can obtain a very competitive performance with a difference less than 0.5\% in Dice to the top performing method as shown in Table~\ref{tab:test}.

\begin{table}[h]
\caption{\label{tab:test} Comparison of segmentation accuracy on the test dataset. Results are from the challenge website (accessed on February 28, 2019).}
\centering
\begin{tabular}{ lccc }
\hline
Methods & \# of Steps & Average Dice (\%) & Global Dice (\%)\\
 \hline
Vorontsov et al. \cite{8363817} & 1 & 95.1    & -  \\
H-DenseUNet \cite{8379359}   & 2 & 96.1    & 96.5 \\
DeepX \cite{yuan2017hierarchical} & 2 & 96.3  & 96.7  \\
2D DenseUNet \cite{8379359}   & 2 & 95.3    & 95.9 \\
 MIMO-FAN (proposed)   & 1 & 95.8 & 96.2 \\
 \hline
\end{tabular}
\end{table}

\subsection{Ablation study} 
\begin{figure}[h]    
	\center{\includegraphics[width = \textwidth] {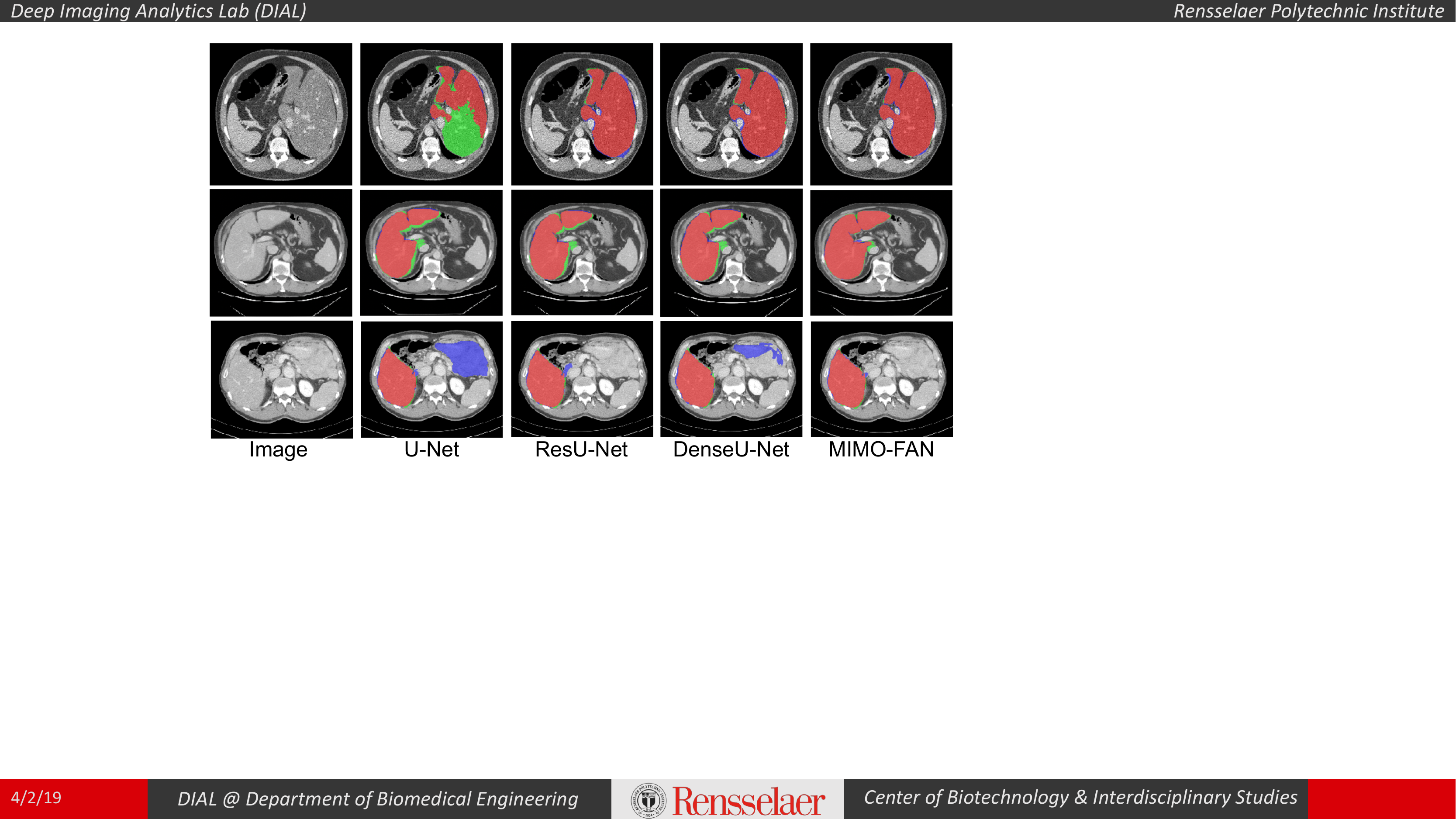}}        
	\caption{Segmentation examples of different methods. From left to right are the raw image, results of U-Net, ResU-Net, DenseU-Net and our proposed MIMO-FAN, the red depicts correctly predicted liver segmentation, the blue shows false positive, green shows false negative.}  
	\label{fig:examples}
\end{figure}

\setlength{\tabcolsep}{3pt}
We further compared our proposed MIMO-FAN against several other classical networks, including U-Net, ResU-Net, and DenseU-Net, to demonstrate the effectiveness of DCC and DPS. Some example results are shown in Fig.~\ref{fig:examples}. The U-Net, ResU-Net and MIMO-FAN are all 19-layer networks with the same numbers of filters. We train all these 2D networks from scratch in the same environment. 
The five-fold cross validation results are shown in Table~\ref{tab:val}.
The conducted $t$-test shows that MIMO-FAN significantly outperforms U-Net, ResU-Net and DenseU-Net with $p$-values of 0.004, 0.025, and 0.002, respectively. 

\begin{table}[h]
	\caption{\label{tab:val} Network ablation study using five-fold cross validation (Dice \%)}
	\centering
	\begin{tabular}{lcccccc}
		\hline
		
		Architecture & Fold 1 & Fold 2 & Fold 3 & Fold 4 & Fold 5 &  Mean$\pm$std\\
		\hline
		U-Net \cite{unet}   & 94.5 & 93.8  & 94.1 & 93.0 & 94.1 & 93.9 $\pm$ 0.50 \\
		ResU-Net \cite{han_automatic_2017} & 94.5 & 94.1  & 94.9 & 92.4 & 94.5 & 94.1 $\pm$ 0.88\\
		DenseU-Net \cite{8379359} &  94.1  & 94.2 & 93.9 & 93.6 &94.5 & 94.1 $\pm$ 0.30\\
		\hline
		MIMO-FAN (DCC) & 95.2 & 93.8 & 94.1 & 92.7 & 94.2 & 94.0 $\pm$ 0.80\\
		MIMO-FAN (DPS) & 95.7 & 94.3 & \textbf{95.0} & 94.6 & 96.1 & 95.1 $\pm$ 0.67 \\
		MIMO-FAN (DCC+DPS) & 96.0 & 95.3 & 94.3 & 95.4 & 96.1 & 95.4 $\pm$ 0.64\\
		MIMO-FAN (DCC+DPS+SF) &\textbf{96.2} & \textbf{95.6} & 94.6 & \textbf{95.7} & \textbf{96.2} & \textbf{95.7} $\pm$ \textbf{0.59}\\
		\hline
	\end{tabular}
\end{table}

\section{New or breakthrough work to be presented}
To the best of our knowledge, the proposed MIMO-FAN is the first network architecture that integrates multi-scale input and multi-scale output features in one single network for efficient abstraction \footnote{The work hasn't been submitted for publications or presentations elsewhere}. We have extensively evaluated the method on the dataset of MICCAI 2017 Liver Tumor Segmentation (LiTS) Challenge\footnote{https://competitions.codalab.org/competitions/17094} dataset and demonstrated promising performance. 

\section{Conclusion}

In this paper, we propose a novel network architecture for unified multi-scale feature abstraction, which incorporates multi-scale features in a hierarchical fashion at various depths for image segmentation. 
The 2D network shows very competitive performance compared with other 3D networks in liver CT image segmentation with a single step. The proposed method can also be applied to other segmentation tasks and will be tested in our future work.

\bibliographystyle{spiebib}
\bibliography{ref} 

\end{document}